\documentclass[conference]{IEEEtran}
\usepackage{subfigure}
\usepackage{color}
\usepackage{amsmath}
\usepackage{cite}
\usepackage{graphicx}
\usepackage{epstopdf}
\usepackage{amsfonts,amsmath,amssymb}
\usepackage{graphicx}
\usepackage{url}
\usepackage{bm}
\usepackage{bbm}
\usepackage{subfigure}
\usepackage{stfloats}

\usepackage{cite,graphicx,amsmath,amssymb,cite,algorithm}

\begin{document}
\newtheorem{theorem}{\textbf{Theorem}}
\newtheorem{proposition}{\textbf{Proposition}}
\newtheorem{corollary}{\textbf{Corollary}}
\newtheorem{remark}{Remark}

\title{Directional Modulation-Enabled Secure Transmission with Intelligent Reflecting Surface}
\author{\IEEEauthorblockN{Liangling Lai$^{\ast}$, Jinsong Hu$^{\ast}$, Youjia Chen$^{\ast}$, Haifeng Zheng$^{\ast}$, and Nan Yang$^{\dag}$}
\IEEEauthorblockA{$^{\ast}$College of Physics and Information Engineering, Fuzhou University, Fuzhou, China}
\IEEEauthorblockA{$^{\dag}$Research School of Electrical, Energy and Materials Engineering, Australian National University, Canberra, Australia}
\IEEEauthorblockA{Emails: lailiangling@gmail.com, \{jinsong.hu, youjia.chen, zhenghf\}@fzu.edu.cn, nan.yang@anu.edu.au}}

\maketitle

\begin{abstract}
We propose a new secure transmission scheme which uses directional modulation (DM) with artificial noise and is aided by the intelligent reflecting surface (IRS). Specifically, the direct path and IRS-enabled reflect path carry the same confidential signal and thus can be coherently added at the desired position to maximize the total received power, while the received signals at other positions are distorted. We derive a closed-form expression for the secrecy rate achieved by the proposed scheme. Using simulation results, we show that the proposed scheme can achieve two-dimensional secure transmission at a specific position. Also, its performance advantage over the conventional DM scheme becomes more pronounced as the number of reflecting elements at the IRS increases.
\end{abstract}


\section{Introduction}\label{sec:introduction}

Intelligent reflecting surface (IRS) has recently emerged as a promising and cutting-edge technology for the fifth generation (5G) and beyond wireless networks, which aims to improve  signal coverage and quality by controlling the low-cost passive reflecting elements integrated on a surface \cite{Renzo2019,Qingqing2020,Bjornson2020,Wu2019}. Specifically, an IRS comprises a huge number of low-cost passive reflecting elements, which can adjust the phase shift of the incident wave, provide favourable wireless communication channels, and reduce the radio waves emission at the transmitter. It is noted that 5G networks may operate at high frequencies (e.g., 4.5--6 GHz). This would incur transmission outage since signals can be easily blocked by moving humans and static obstacles in an enclosed room. This blockage problem can be addressed by using the IRS to control the propagation environment, since the blocked signal can be reflected by the IRS to the desired user. Thus, IRS is a compelling wireless technology, particularly for indoor applications with the high density of users (e.g., stadium, shopping mall, and exhibition center).

In wireless communication systems, physical layer security is a fundamental but formerly untapped solution to secrecy problems, which aims to guarantee the confidentiality of transmitted information \cite{Yang2015}. Comparing to the techniques using higher-layer encryption, physical layer security explores the characteristics of the communications medium to increase the signal error of the eavesdropper. Among various techniques, directional modulation (DM) is one of the effective schemes to enhance physical layer security \cite{Ding2014,Hu2016,Hu2017,Feng2018}. Specifically, DM can distort the constellation of the signals received at potential eavesdroppers while retaining the original shape of the signals received at the legitimate user. As indicated in \cite{Ding2014,Hu2016}, the signal transmission in the DM scheme depends delicately on the line-of-sight (LoS) path, since the unwanted artificial noise (AN) is introduced towards the desired user in the multipath environment. Motivated by this, we apply the IRS to DM with AN to reinforce the control of the wireless propagation environment and utilize the multipath environment to enhance the physical layer security.

In this work, we propose a novel transmission scheme to enhance the physical layer secrecy performance of the system where a transmitter communicates with its legitimate receiver in the presence of an eavesdropper. The main contributions of this work are summarized as follows.
\begin{itemize}
\item The scheme uses DM with AN and is aided by the IRS, which reinforces the control of the propagation environment and utilizes the static multipath environment to enhance the physical layer security.

\item A closed-form expression is derived for the secrecy rate achieved by the proposed scheme. We also compare the signal-to-noise ratio (SNR) of the proposed scheme with the conventional scheme.

\item Using simulation results, we find that the proposed scheme can achieve two-dimensional (2D) secure transmission. We also find that the proposed scheme outperforms the conventional DM scheme, specifically when more reflecting elements are integrated at the IRS.
\end{itemize}

The remainder of this paper is organized as follows. Section \ref{sec:system_model} details our system model and assumptions. A thorough performance analysis of the proposed scheme is provided in Section \ref{sec:analysis}. Section \ref{sec:simulation} provides simulation results to validate our analysis and provide useful insights into the impact of system
parameters. Section \ref{sec:conclusion} draws the conclusions.

\emph{Notations:} Scalar variables are denoted by italic symbols. Vectors and matrices are denoted by lower-case and upper-case boldface symbols, respectively. Given a complex number, $|\cdot|$ denote the modulus. Given a complex vector or matrix, $(\cdot)^{\ast}$, $(\cdot)^T$, $(\cdot)^H$, and $\|\cdot\|$ denote the conjugate,  transpose, conjugate transpose, and norm, respectively. $\mathbb{C}^{p \times q}$ denotes the space of $p \times q$ complex-valued matrices. The $N \times N$ identity matrix is referred to as $\mathbf{I}_{N}$ and $\mathbb{E}[\cdot]$ denotes expectation operation. $\mathcal{C} \mathcal{N}\left(\mu, \sigma^{2}\right)$ denotes the distribution of a circularly symmetric complex Gaussian (CSCG) random variable with mean $\mu$ and variance $\sigma^{2}$.

\section{System Model}\label{sec:system_model}

\begin{figure}
    \begin{center}
        \includegraphics[width=3.4in, height=3.1in]{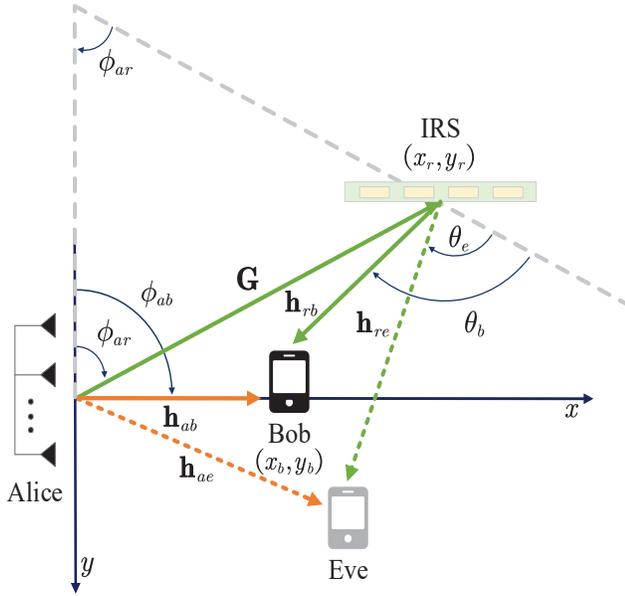}
        \caption{Illustration of the IRS-aided DM with AN transmission.}\label{fig:system_model}\vspace{0mm}
    \end{center}
\end{figure}

As shown in Fig.~\ref{fig:system_model}, we propose an IRS-aided DM wireless transmission scheme. In this scheme, there are two paths from the transmitter, Alice, to the legitimate receiver, Bob. One is the direct Alice-Bob path and the other one is the Alice-IRS-Bob path where the IRS reflects the incident wave from Alice towards Bob. The IRS is equipped with a controller to manipulate the phase of the transmit signal for a better quality of communication. We assume that the locations of Bob and IRS are known by Alice, denoted by $(x_b, y_b)$ and $(x_r, y_r)$, respectively. We also assume that Bob and the eavesdropper, Eve, are equipped with a single antenna.

In the Alice-Bob path, the normalized steering vector from Alice to Bob is given by
\begin{equation}\label{eq:hd}
\mathbf{h}_{ab}=\frac{1}{\sqrt{N_a}}\left[e^{j 2 \pi \Phi_0(\phi_{ab})}, \ldots, 1,  \ldots,  e^{j 2 \pi \Phi_{N_a-1}(\phi_{ab})}\right]^H,
\end{equation}
where 
the phase shift $\Phi_n(\phi)$ is defined as
\begin{equation}\label{eq:Phi}
\Phi_n(\phi)\!=-\frac{d}{\lambda}\left(n\!-\!\frac{N\!-\!1}{2}\right) \cos \phi, ~~n \!=\! 0,1,\ldots,N\!-\!1.
\end{equation}
Here, $\lambda$ is wavelength of the carrier, $N\in\{N_{a},N_{r}\}$ is the number of antennas at Alice or reflecting elements at the IRS, $d\in\{d_a, d_r\}$ is the spacing between the antennas at Alice or the reflecting elements at the IRS, and $\phi\in\{\phi_{ab}, \phi_{ar}\}$ is the transmit angle from Alice to Bob or from Alice to IRS.

In the Alice-IRS-Bob path, the spatial steering matrix from Alice to IRS is given by 
\begin{equation}
\mathbf{G}=\mathbf{g}_r^{\ast}\mathbf{g}_t^H,
\end{equation}
where 
$\mathbf{g}_t$ is the transmit spatial steering vector given by
\begin{equation}\label{gt}
\mathbf{g}_{t}=\frac{1}{\sqrt{N_a}}
\left[e^{j2\pi\Phi_0(\phi_{ar})},\ldots,1,\ldots,e^{j2\pi\Phi_{N_a-1}(\phi_{ar})}\right]^H,
\end{equation}
and $\mathbf{g}_{r} \in \mathbb{C}^{N_r \times 1}$ is the receive steering vector with all-one elements. The IRS can be treated as a reflector with a large number of low-cost passive elements. Thus, it can reflect the electromagnetic waves by adding an additional phase to the received signal. The steering vector from the IRS to Bob and Eve are denoted by $\mathbf{h}_{rb}^H\in \mathbb{C}^{1 \times N_r}$ and $\mathbf{h}_{re}^H\in \mathbb{C}^{1 \times N_r}$, respectively, with all-one elements. The diagonal phase shifting matrix of the IRS is given by
\begin{equation}\label{theta}
\mathbf{\Theta}(\theta)=\operatorname{diag}\left(e^{-j 2 \pi \Phi_0(\theta-\theta_b)}, \cdots, e^{-j 2 \pi \Phi_{N_r-1}(\theta-\theta_b)}\right),
\end{equation}
where $\theta\in\{\theta_b,\theta_e\}$ is the deflection angle for the signal of Bob or Eve at the IRS.

The signal received at Bob from both the Alice-Bob and the Alice-IRS-Bob paths can be expressed as
\begin{equation}\label{eq:y_b}
y_{b}=\sqrt{P_t}\mathbf{H}_b\mathbf{x}+n_b,
\end{equation}
where $n_b\sim\mathcal{CN}(0,\sigma^2)$ is the additive white Gaussian noise (AWGN) at Bob, ${P_t}$ is the transmit power. The complex baseband transmitted signal at Alice is given by
\begin{equation}
\mathbf{x}=\left[\mathbf{x}_a~\mathbf{x}_r\right]^{T}.
\end{equation}
Moreover, $\mathbf{H}_b$ denotes the two paths from Alice to Bob, given by
\begin{equation}
\mathbf{H}_b=\left[\sqrt{L_{ab}}\mathbf{h}_{ab}^H~~ \sqrt{L_{arb}}\mathbf{h}_{rb}^{H}\mathbf{\Theta}(\theta_b)\mathbf{G}\right]. 
\end{equation}
Due to the time difference between the two paths, we assume that the confidential signals are carefully designed to transmit at different time slots, allowing Bob to receive the signals from two paths at the same time. Hence, we have $\mathbf{x}_a=\sqrt{\alpha}\mathbf{w}_a s+\sqrt{1-\alpha}\mathbf{P}_a \mathbf{z}$ and $\mathbf{x}_r=\sqrt{\alpha}\mathbf{w}_r s$, where $s$ denotes the confidential data for Bob with the average power $\mathbb{E}\left[|s|^{2}\right]=1$, $\mathbf{w}_a$ and $\mathbf{w}_r$ denote the precoding vectors of the Alice-Bob path and the Alice-IRS-Bob path, respectively.
Moreover, we have $\mathbf{P}_a \triangleq\left(\mathbf{I}_{N_a}-\mathbf{h}_{ab}\mathbf{h}_{ab}^{H}\right)/
\|\mathbf{I}_{N_a}-\mathbf{h}_{ab}\mathbf{h}_{ab}^{H}\|$ as the vector to transmit AN signals towards Eve \cite{Hu2016}, where
$\mathbf{z}$ consists of $N_a$ i.i.d. circularly-symmetric complex Gaussian random variables with zero-mean and unit-variance, i.e., $\mathbf{z}\sim\mathcal{CN}(0,\mathbf{I}_{N_a})$, and $\alpha$ denotes the power allocation factor between useful signals and AN signals. We next adopt $L_{x y}=\left(d_{x y}/d_{0}\right)^{-2}$ as the free-space path loss model, where $d_0$ is the reference distance and $d_{x y}$ is the distance between $x$ and $y$. Accordingly, $L_{ab}$ denotes the Alice-Bob distance and $L_{arb}$ denotes the Alice-IRS-Bob distance. In order to maximize the SNR at Bob, we set $\mathbf{w}_a$ and $\mathbf{w}_r$ as $\mathbf{w}_a=\mathbf{h}_{ab}$ and $\mathbf{w}_r=\mathbf{g}_{t}$, respectively.

The received signal at Eve is expressed
\begin{equation}
y_{e}=\sqrt{P_t}\mathbf{H}_{e}\mathbf{x}+n_e,
\end{equation}
where $n_e\sim\mathcal{CN}(0,\sigma^2)$ is the AWGN at Eve and
$\mathbf{H}_e$ denotes the two paths from Alice to Eve, given by
\begin{equation}
\mathbf{H}_e=\left[\sqrt{L_{ae}}\mathbf{h}_{ae}^{H}~~ \sqrt{L_{are}}\mathbf{h}_{re}^{H}\mathbf{\Theta}(\theta_e)\mathbf{G}\right].
\end{equation}

In order to characterize the performance at Bob, we calculate the bit error rate (BER) as
\begin{equation}
\mathrm{BER}=\frac{2}{\log_{2}M}Q\left(\sqrt{2\gamma}\sin\left(\frac{\pi}{M}\right)\right),
\end{equation}
where $\gamma$ is the SNR or the signal-to-interference-plus-noise ratio (SINR) at receiver, and the $Q$-function is defined as $Q(u)=1/\sqrt{2\pi}\int_{u}^{\infty}e^{-u^{2}/2}\mathrm{d}u$. In this work, we consider the use of quadrature phase shift keying (QPSK) such that the BER is given by $\textrm{BER}=Q\left(\sqrt{\gamma}\right)$.

\section{Secrecy Performance Analysis}\label{sec:analysis}

In this section, we analyze the secrecy performance of the proposed scheme.By using the expressions for $\mathbf{x}_a$ and $\mathbf{x}_r$ and noting that $\mathbf{h}_{ab}^{H}\mathbf{w}_a=1$, the received signal at Bob is rewritten as
\begin{align}\label{eq:y_b1}
y_{b}=&\sqrt{P_t{L_{ab}}}\mathbf{h}_{ab}^{H}\left(\sqrt{\alpha}\mathbf{w}_a s+\sqrt{1-\alpha}\mathbf{P}_a\mathbf{z}\right)\notag\\
&+\sqrt{P_t{L_{arb}}}\mathbf{h}_{rb}^H \mathbf{\Theta}(\theta_b)\mathbf{G}(\sqrt{\alpha}\mathbf{w}_r s)+n_b\notag\\
=&\left(\sqrt{\alpha P_t{L_{ab}}}+\sqrt{\alpha P_t{L_{arb}}}\mathbf{h}_{rb}^{H} \mathbf{\Theta}(\theta_b)\mathbf{G} \mathbf{w}_r \right)s+n_b,
\end{align}
where $\mathbf{P}_a\mathbf{z}$ is orthogonal to $\mathbf{h}_{ab}^H$, implying $\mathbf{h}_{ab}^H\mathbf{P}_a\mathbf{z}=0$. As per \eqref{gt} and \eqref{theta}, we obtain
\begin{align} \label{eq:hr*G*omega}
\mathbf{h}_{rb}^{H} \mathbf{\Theta}(\theta_b) \mathbf{G} \mathbf{w}_r
&=\frac{1}{N_a}\sum_{k=0}^{N_a-1}\sum_{l=0}^{N_r-1}e^{j2\pi\left(\psi_{1}+\psi_{2}\right)}\\
&=N_r,
\end{align}
where $\psi_1=\Phi_k(\phi_{ar})-\Phi_k(\phi_{ar})$ and $\psi_2=\Phi_l(\theta_b)-\Phi_l(\theta_b)$. Therefore, the SNR at Bob is given by
\begin{equation}\label{eq:SNRbob}
\gamma_{b}=\frac{\alpha P_t\left|\sqrt{L_{ab}}+\sqrt{L_{arb}}N_r \right|^{2}}{\sigma^{2}}.
\end{equation}

Similarly, the received signal at Eve is rewritten as
\begin{align}
y_{e}=&\sqrt{P_t{L_{ae}}}\mathbf{h}_{ae}^{H}\left(\sqrt{\alpha}\mathbf{w}_a s+\sqrt{1-\alpha}\mathbf{P}_a \mathbf{z}\right)\notag\\
&+\sqrt{P_t{L_{are}}}\mathbf{h}_{re}^H\mathbf{\Theta}\left(\theta_e\right)
\mathbf{G}\left(\sqrt{\alpha}\mathbf{w}_r s\right)+n_e\notag\\
=&\left(\sqrt{\alpha P_t{L_{ae}}}\mathbf{h}_{ae}^{H}\mathbf{w}_a+\sqrt{\alpha P_t{L_{are}}}\mathbf{h}_{re}^{H}\mathbf{\Theta}\left(\theta_e\right)
\mathbf{G}\mathbf{w}_r\right)s\notag\\
&+\sqrt{\left(1-\alpha\right)P_t{L_{ae}}}\mathbf{h}_{ae}^{H}\mathbf{P}_a\mathbf{z}+n_e.
\end{align}
Accordingly, the signal-to-interference-plus-noise ratio (SINR) at Eve can be given by
\begin{equation}\label{eq:SINReve}
\gamma_{e}=\frac{\alpha P_t\left|\sqrt{L_{ae}}\mathbf{h}_{ae}^{H} \mathbf{w}_a+\sqrt{L_{are}}\mathbf{h}_{re}^{H}\mathbf{\Theta}\left(\theta_e\right)\mathbf{G} \mathbf{w}_r\right|^{2}}{(1-\alpha)P_t\left|\sqrt{L_{ae}}\mathbf{h}_{ae}^{H}\mathbf{P}_a \mathbf{z}\right|^2+\sigma^{2}},
\end{equation}
where
\begin{align}\label{eve}
\mathbf{h}_{re}^{H}\mathbf{\Theta}\left(\theta_e\right)\mathbf{G}\mathbf{w}_r
=&\frac{1}{N_a}\sum_{k=0}^{N_a-1}\sum_{l=0}^{N_r-1}e^{j2\pi
\left(\psi_{1}+\Phi_l\left(\theta_e\right)-\Phi_l\left(\theta_b\right)\right)}\notag\\
=&\frac{e^{j \pi\frac{N_r}{2}\left(\cos\theta_e-\cos\theta_b\right)}
-e^{-j\pi\frac{N_r}{2}\left(\cos\theta_e-\cos\theta_b\right)}}
{e^{j\pi\frac{1}{2}\left(\cos\theta_e-\cos\theta_b\right)}
-e^{-j\pi\frac{1}{2}\left(\cos\theta_e-\cos\theta_b\right)}}\notag\\
=&\frac{\sin\left(\frac{N_r\pi}{2}\left(\cos\theta_e-\cos\theta_b\right)\right)}
{\sin\left(\frac{\pi}{2}\left(\cos\theta_e-\cos\theta_b\right)\right)}\notag\\
\leq& N_r.
\end{align}
Following \eqref{eve} and noting that $\mathbf{h}_{ae}^{H} \mathbf{w}_a\leq 1$ and $\mathbf{h}_{ae}^{H}\mathbf{P}_a \mathbf{z}\geq0$, we find that $\gamma_{e}$ is lower than $\gamma_{b}$.

In the considered system, the rate at Bob and Eve are given by $R_{b}=\log_2\left(1+\gamma_{b}\right)$ and $R_{e}=\log_2\left(1+\gamma_{e}\right)$, respectively. Thus, as per the rules of physical layer security, the secrecy rate of the proposed scheme is calculated as
\begin{equation}
R_{s}=\left[R_{b}-R_{e}\right]^{+}=\log_2 \left(\frac{1+\gamma_{b}}{1+\gamma_{e}}\right),
\end{equation}
where $[x]^{+}\triangleq\max \{0, x\}$. Since we find $\gamma_{e}<\gamma_{b}$, $R_{s}$ is always a positive value.

\section{Simulation Results and Discussion}\label{sec:simulation}

In this section, we use simulations to evaluate the secrecy performance of the proposed scheme. Unless stated otherwise, we consider  the number of antenna at Alice is set as $N_a=16$ and the number of elements of IRS is set as $N_r=50$. The noise power is set as $\sigma^2=-20~\textrm{dBm}$ and transmit power $P_t=25~\textrm{dBm}$. The reference distance is $d_0=1~\textrm{m}$ and power allocation factor between useful signals and AN signals is $\alpha=0.6$. The spacing between antennas or reflecting elements is half of the wavelength. The locations of Alice, Bob, and IRS are set as $\mathrm{Alice}\left(0,0\right)$, $\mathrm{Bob}\left(20,0\right)$, and $\mathrm{IRS}\left(20,-15\right)$, respectively.

\begin{figure}[t!]
    \begin{center}
    \includegraphics[width=3.3in,height=2.9in]{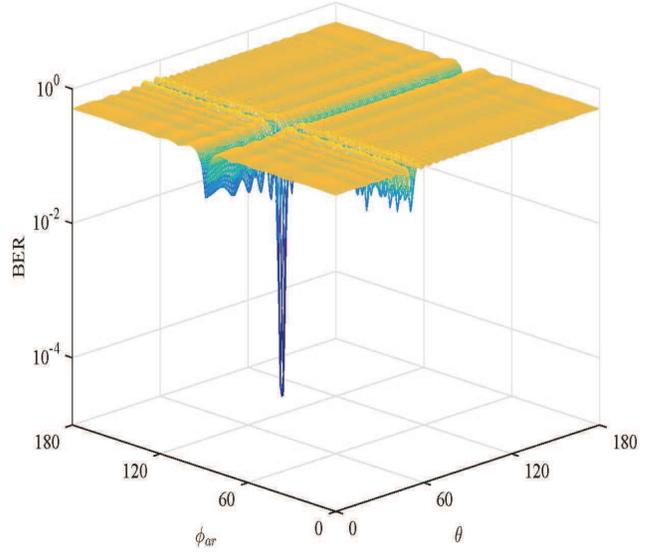}
    \caption{The BER of our considered system versus $\phi_{ar}$ and $\theta$.}\label{fig:2ps}
    \end{center}
\end{figure}

Fig.~\ref{fig:2ps} shows the BER of our considered system versus $\phi_{ar}$ and $\theta$ ranging from $0^{\circ}$ to $180^{\circ}$. We observe from this figure that there are two lines achieving the lower BER than other lines. Specifically, one line is the Alice-Bob path and the other line is the Alice-IRS-Bob path. Interestingly, the intersection of these two paths, which achieves the minimum BER by our proposed scheme, is the location of Bob, which is our desired direction. We also observe that the BER rapidly becomes worse in undesired directions, e.g. the BER hovers around 0.5. This is due to the fact that the signal along the desired direction is enhanced by the Alice-IRS-Bob path but not impacted by AN signals, while the signals along the undesired directions are severely impacted by AN signals. This indicates that our proposed scheme achieves 2D secure transmission by taking the advantage of the additional path enabled by the IRS.

\begin{figure}[t!]
    \begin{center}
    \includegraphics[width=3.5in,height=2.9in]{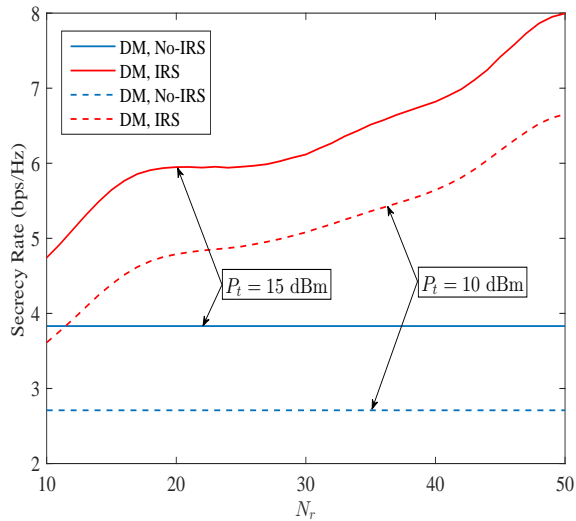}
    \caption{The secrecy rate versus $N_r$ for different values of $P_{t}$.}
    \label{fig:3ps}
    \end{center}
   \end{figure}

In Fig.~\ref{fig:3ps}, we compare the secrecy performance of our proposed scheme, referred to as ``DM, IRS'', with that of the benchmark scheme where DM is applied without the IRS, referred to as ``DM, No-IRS''. The secrecy rate, $R_{s}$, achieved by the two schemes are plotted versus $N_r$ for $P_t=10~\textrm{dBm}$ and $P_t=15~\textrm{dBm}$. We observe that for our proposed scheme, $R_{s}$ increases when $N_r$ increases, while for the benchmark scheme, $R_{s}$ remains unchanged. We then observe that the secrecy performance advantage of our proposed scheme over the benchmark scheme becomes more profound when $N_r$ increases. These observations indicate that using the IRS can effectively enhance the security performance of a wireless communication system with low power consumption.

\begin{figure}[t!]
     \begin{center}
    \includegraphics[width=3.5in,height=2.9in]{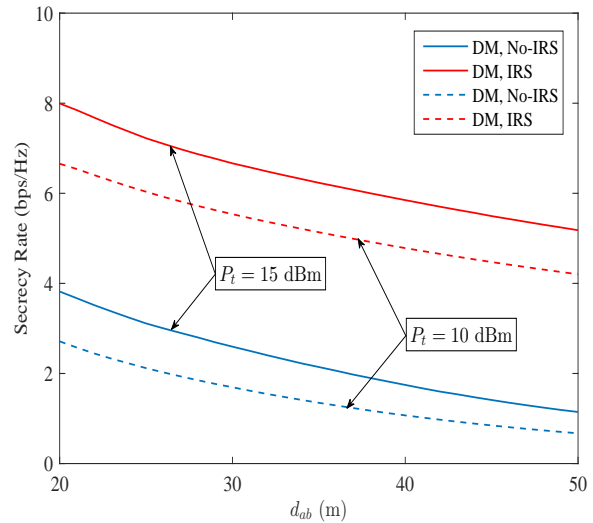}
    \caption{The secrecy rate versus $d_{ab}$ under different values of ${P_t}$.}
    \label{fig:4ps}
    \end{center}
\end{figure}

Fig.~\ref{fig:4ps} shows the secrecy rate $R_{s}$ versus the distance between Alice and Bob, $d_{ab}$, for $P_t=10~\textrm{dBm}$ and $P_t=15~\textrm{dBm}$. We first observe that our proposed scheme always outperform the benchmark scheme, by achieving a higher secrecy rate, when $d_{ab}$ increases. Second, we observe that the secrecy rate achieved by both schemes decreases when $d_{ab}$ increases. Third, we observe that when $d_{ab}$ is sufficient large (e.g., $d_{ab}=50~$m), the secrecy performance advantage of our proposed scheme over the benchmark scheme for $P_t=15~\textrm{dBm}$ is more significant than that for $P_t=10~\textrm{dBm}$. This is mainly due to the fact that the transmit power constraint at Alice is a main contributor to the secure transmission aided by the passive IRS.

\section{Conclusion}\label{sec:conclusion}

A new IRS-aided DM with AN scheme was proposed in this work to utilize the multipath propagation environment for enhancing the physical layer security. In order to examine the secrecy performance of the proposed scheme, we derived a closed-form expression for the secrecy rate. Simulation results showed that 2D secure transmission can be achieved by our proposed scheme. We found that our proposed scheme outperforms the conventional DM scheme which does not use IRS, and this performance advantage increases when there are more reflecting elements at the IRS. Additionally, we found that the transmit power constraint at the transmitter plays a key role in governing the secrecy rate improvement of our proposed scheme when the distance between the transmitter and the legitimate receiver is sufficiently large.

\bibliographystyle{IEEEtran}
\bibliography{references}

\begin{thebibliography}{1}
\providecommand{\url}[1]{#1}
\csname url@samestyle\endcsname
\providecommand{\newblock}{\relax}
\providecommand{\bibinfo}[2]{#2}
\providecommand{\BIBentrySTDinterwordspacing}{\spaceskip=0pt\relax}
\providecommand{\BIBentryALTinterwordstretchfactor}{4}
\providecommand{\BIBentryALTinterwordspacing}{\spaceskip=\fontdimen2\font plus
\BIBentryALTinterwordstretchfactor\fontdimen3\font minus
  \fontdimen4\font\relax}
\providecommand{\BIBforeignlanguage}[2]{{%
\expandafter\ifx\csname l@#1\endcsname\relax
\typeout{** WARNING: IEEEtran.bst: No hyphenation pattern has been}%
\typeout{** loaded for the language `#1'. Using the pattern for}%
\typeout{** the default language instead.}%
\else
\language=\csname l@#1\endcsname
\fi
#2}}
\providecommand{\BIBdecl}{\relax}
\BIBdecl

\bibitem{Renzo2019}
M.~D. Renzo, M.~Debbah, and D.~T. {Phan-Huy et al.}, ``{Smart radio
  environments empowered by reconfigurable AI meta-surfaces: an idea whose time
  has come},'' \emph{EURASIP J. Wireless Commun. Netw}, vol. 2019, no.~1, May
  2019.

\bibitem{Qingqing2020}
Q.~Wu and R.~Zhang, ``Towards smart and reconfigurable environment: Intelligent
  reflecting surface aided wireless networks,'' \emph{IEEE Commun. Mag.},
  vol.~58, no.~1, pp. 106--112, Jan. 2020.

\bibitem{Bjornson2020}
O.~Ozdogan, E.~Bjornson, and E.~G. Larsson, ``Intelligent reflecting surfaces:
  Physics, propagation, and pathloss modeling,'' \emph{IEEE Wireless Commun.
  Lett.}, vol.~9, no.~5, pp. 581--585, May 2020.

\bibitem{Wu2019}
Q.~Wu and R.~Zhang, ``Intelligent reflecting surface enhanced wireless network
  via joint active and passive beamforming,'' \emph{IEEE Trans. Wireless
  Commun.}, vol.~18, no.~11, pp. 5394--5409, Nov. 2019.

\bibitem{Yang2015}
N.~Yang, L.~Wang, G.~Geraci, M.~Elkashlan, J.~Yuan, and M.~{Di Renzo},
  ``{Safeguarding 5G wireless communication networks using physical layer
  security},'' \emph{IEEE Commun. Mag.}, vol.~53, no.~4, pp. 20--27, Apr. 2015.

\bibitem{Ding2014}
Y.~Ding and V.~Fusco, ``A vector approach for the analysis and synthesis of
  directional modulation transmitters,'' \emph{IEEE Trans. Antennas Propag.},
  vol.~62, no.~1, pp. 361--370, Jan. 2014.

\bibitem{Hu2016}
J.~Hu, F.~Shu, and J.~Li, ``Robust synthesis method for secure directional
  modulation with imperfect direction angle,'' \emph{IEEE Commun. Lett.},
  vol.~20, no.~6, pp. 1084--1087, Jun. 2016.

\bibitem{Hu2017}
J.~Hu, S.~Yan, F.~Shu, J.~Wang, J.~Li, and Y.~Zhang, ``Artificial-noise-aided
  secure transmission with directional modulation based on random frequency
  diverse arrays,'' \emph{IEEE Access}, vol.~5, pp. 1658--1667, Jan. 2017.

\bibitem{Feng2018}
F.~Shu, X.~Wu, J.~Hu, J.~Li, R.~Chen, and J.~Wang, ``Secure and precise
  wireless transmission for random-subcarrier-selection-based directional
  modulation transmit antenna array,'' \emph{IEEE J. Sel. Areas Commun.},
  vol.~36, no.~4, pp. 890--904, Apr. 2018.

\end{thebibliography}

\end{document}